\documentclass[aps,pra,twocolumn,groupedaddress,reprint,showpacs]{revtex4-2}
\usepackage{graphicx}
\usepackage{setspace}
\usepackage{amsmath}
\usepackage{amssymb}
\usepackage{color}
\usepackage{array}
\usepackage{subfigure}
\usepackage{hyperref}
\usepackage{float}
\usepackage{lipsum}

\usepackage[all]{xy}
\newcommand*{\Scale}[2][4]{\scalebox{#1}{$#2$}}%
\newcommand{\RN}[1]{%
  \textup{\uppercase\expandafter{\romannumeral#1}}%
}

\begin{document}

\title{Design fast Rydberg blockade SWAP gates with synthetic modulated driving}
\author{Xin Wang}
\author{Tianze Sheng}
\author{Yuan Sun}
\email[email: ]{yuansun@siom.ac.cn}
\affiliation{CAS Key Laboratory of Quantum Optics and Center of Cold Atom Physics, Shanghai Institute of Optics and Fine Mechanics, Chinese Academy of Sciences, Shanghai 201800, China}
\affiliation{University of Chinese Academy of Sciences, Beijing 100049, China}

\begin{abstract}
The cold atom qubit platform emerges as an attractive choice for the next stage of quantum computation research, where a special family of synthetic analytical pulses has considerably improved the experimental performance of Controlled-PHASE Rydberg blockade gates in recent studies. The success of Controlled-PHASE Rydberg blockade gates triggers the intriguing question of whether the two-qubit Rydberg blockade gate SWAP gate exists. Via investigating the transition linkage structure, we provide a definitive answer to this question and establish the method of fast SWAP Rydberg blockade gates with synthetic continuously-modulated driving. These gate protocols use careful analysis to properly generate coherent population transfer and phase accumulation of the wave function in the atom-laser interaction process. They can adapt to finite Rydberg blockade strengths and bear considerable resistance some major adverse effects such as laser fluctuations. Further examinations reveal that we can anticipate satisfying performances of the method with currently available experimental techniques in relevant research areas.
\end{abstract}
\maketitle

\section{Introduction}

The cold atom qubit platform \cite{PhysRevA.62.052302, RevModPhys.82.2313, J.Phys.B.49.202001} has been evolving rapidly since the first experimental demonstrations of Rydberg blockade gate \cite{nphys1178, PhysRevLett.104.010503}. Recently, the ground-Rydberg excitation laser has improved significantly \cite{PhysRevA.97.053803, YouLi2024Optica}, the coherent quantum control of Rydberg atoms has been making steady progress \cite{nphys3487, Saffman2019prl, PhysRevApplied.15.054020, Kaufman2022nphys}, and shuttling atoms to form defect-free array has become a helpful tool \cite{Ahn2016NC, science.aah3778, science.aah3752}. Most interestingly, on the subject of the two-qubit Controlled-PHASE gate as the essential ingredient of quantum computation hardwares, the Rydberg blockade gate has received substantial advancement in fidelity \cite{PhysRevA.105.042430, Lukin2023nature}, thanks to the method of continuously-modulated off-resonant pulse replacing the early proposal of discrete resonant pulse sequence \cite{PhysRevLett.85.2208, PhysRevApplied.13.024059, PhysRevApplied.15.054020}. Nevertheless, the Rydberg dipole-dipole interaction does not naturally extend to very long ranges \cite{RevModPhys.82.2313, PhysRevA.92.042710, PhysRevA.107.063711}, and therefore enhancing the connectivity of Rydberg blockade Controlled-PHASE gate has to resort to special techniques. Now that the number of cold atom qubits loaded in an array has convincingly surpassed the level of 1000 \cite{Pause2024Optica}, finding a path to realize high-connectivity entangling quantum logic gates for very-large-scale array becomes an imminent challenge. We are looking forward to a realistic solution where qubit atoms can stay stationary during the gate operations such as the buffer-atom-mediated (BAM) gate and the buffer-atom framework \cite{Yuan2024SCPMA}, while mechanically shuttling the qubit atoms across the array seems slow in speed and may cause extra sources of leakage \cite{Saffman2024arXiv}. Like the role of SWAP gate in the superconducting qubit platform \cite{PhysRevApplied.6.064007, PhysRevX.11.021058, FR_FAN20215}, to simultaneously obtain high connectivity, high speed and high fidelity for sizable quantum algorithms, the future cold atom qubit platform is calling for a fast SWAP gate that will eventually work together with the buffer atoms and ancilla atoms \cite{Yuan2024SCPMA}.

With the exciting achievements in the last decade, the Rydberg blockade effect \cite{nphys1178, nphys1183, RevModPhys.82.2313, PhysRevA.92.042710, J.Phys.B.49.202001} has been extensively adopted in various quantum information processing tasks for its robustness and versatility. So far, the Rydberg blockade effect has repeatedly proved its better experimental compatibility compared with other schemes utilizing the Rydberg dipole-dipole interactions such as Rydberg pumping or Rydberg anti-blockade \cite{Wu2021PRJ, PhysRevA.109.012608, EPJQT40507-024-00246} which often require a precisely fixed value of Rydberg dipole-dipole interaction strength. In particular, the recent success of applying synthetic continuously-modulated driving in constructing high-fidelity Rydberg blockade entangling phase gates inspires us to ask the question whether a fast Rydberg blockade SWAP gate is possible via similar methods. 

In this article, we provide a definitive answer that the Rydberg blockade SWAP gate protocol does exist. Moreover, we are going to design the fast Rydberg blockade SWAP gate to meet the current experimental conditions and analyze its principles and characteristics. Not surprisingly, the anticipated gate protocols have close ties with the quantum geometry \cite{BerryPhasePaper, PhysRevLett.58.1593, PhysRevLett.131.240001}. The rest of contents are organized as follows. At first, we introduce the basic mechanisms of fast Rydberg blockade SWAP gate and demonstrate the typical routines of how to design proper gate protocols to satisfy the practical requirements. Subsequently we also discuss the influence of typical adverse effects that reduce the fidelity of Rydberg blockade gates, where we keep adopting the method of suppressing the high-frequency components in the driving waveforms \cite{PhysRevApplied.20.L061002}. While these contents focus on the one-photon ground-Rydberg transition, we observe that the main idea becomes immediately applicable to continuously-modulated pulse in two-photon ground-Rydberg transition \cite{OptEx480513} at almost no extra burden in designing. We will evaluate the two-qubit SWAP gate fidelity according to the well-established criteria \cite{PEDERSEN200747, PhysRevA.96.042306}.

\section{Atom-laser interaction for Rydberg blockade SWAP gate}

\begin{figure}[h]
\centering
\includegraphics[width=0.46\textwidth]{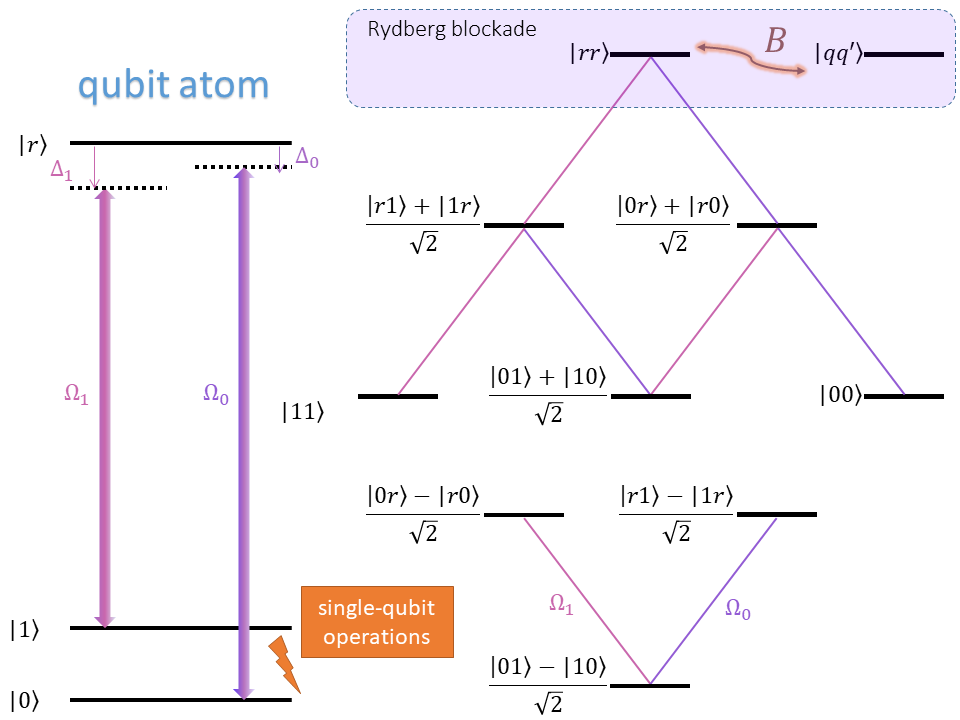}
\caption{(Color online) The optical drivings and atomic transition linkage structure of atom-laser interaction under study. The two driving lasers $\omega_0, \omega_1$ form a $\Lambda$-like transition pattern on a single qubit atom. The detailed transition linkage pattern with respect to the two-qubit basis states is shown in the right, which can also be interpreted by the Morris-Shore transform \cite{PhysRevA.27.906}.}
\label{fig1:layout_sketch}
\end{figure}

We sketch the basic ingredients of atom-laser transitions and the relevant linkage structure in Fig. \ref{fig1:layout_sketch}. Without loss of generality, we choose two non-degenerate magnetically-insensitive states among the ground or metastable ground levels as the two qubit register states, which can be instantiated in typical alkali and alkali earth atoms. In this SWAP gate protocol, we need two mutually coherent lasers $\omega_0, \omega_1$ to drive the ground-Rydberg transition $|0\rangle \leftrightarrow |r\rangle, |1\rangle \leftrightarrow |r\rangle$ respectively. Assume uniform optical drivings on the control and target atoms as the usual practice in experimental implementations now, then the Hamiltonian $H_\text{a}$ for a single control or qubit atom is:
\begin{equation}
\label{eq:Hc_1ph}
H_\text{a} = \frac{1}{2}\hbar\Omega_0 e^{i\Delta_0 t} |0\rangle\langle r| + \frac{1}{2}\hbar\Omega_1 e^{i\Delta_1 t} |1\rangle\langle r| + \text{H.c.},
\end{equation}
where we define the detuning terms as $\omega_\text{transition} - \omega_\text{laser}$ and let the rotating wave frequency be $\omega_\text{transition}$. Both the SWAP gate behavior and the Rydberg blockade effect manifest themselves beyond single-body phenomena, and therefore it seems not only natural but also necessary to carry on the analysis under the two-qubit basis. Besides the single-body Hamiltonians of the control and target qubit atoms in Eq. \eqref{eq:Hc_1ph}, the two-body interaction process also contains the Rydberg blockade effect as modeled by the F\"oster resonance below: 
\begin{equation}
\label{eq:Rydberg_blockade}
H_{q} = B|rr\rangle \langle qq'| + B|qq'\rangle \langle rr| + \delta_q |qq'\rangle\langle qq'|,
\end{equation}
where $B$ is a real number denoting the strength of blockade and $\delta_q$ is the penalty energy term. Then the four orthonormal basis two-qubit states divide themselves into two mutually independent groups of different dynamics: $\frac{1}{\sqrt{2}}(|01\rangle - |10\rangle)$, and $|00\rangle, \frac{1}{\sqrt{2}}(|01\rangle + |10\rangle), |11\rangle$, which we name as singlet and triplet respectively. According to the corresponding linkage structures shown in Fig. \ref{fig1:layout_sketch}, the dynamics of singlet does not involve the Rydberg blockade effect due to symmetry reasons. More details of the Hamiltonian and time evolution calculations are provided in the appendix.

\begin{table}
\begin{center}
\Scale[1.0]{
  \begin{tabular}{ | c | c | }  
    \hline
    input state & output state \\ \hline
    $|00\rangle$ & $|00\rangle$ \\ \hline
    $|01\rangle$ & $|10\rangle$ \\ \hline      
    $|10\rangle$ & $|01\rangle$ \\ \hline  
    $|11\rangle$ & $|11\rangle$ \\ \hline         
  \end{tabular}\,
  \begin{tabular}{ | c | c | }  
    \hline
    input state & output state \\ \hline
    $|10\rangle$ & $|10\rangle$ \\ \hline
    $|01\rangle$ & $|01\rangle$ \\ \hline      
    $|00\rangle$ & $|11\rangle$ \\ \hline  
    $|11\rangle$ & $|00\rangle$ \\ \hline  
  \end{tabular}  
  }
\end{center}
\caption{Two formats of SWAP gate under consideration. They are equivalent up to local single-qubit gates.}
\label{tab:swap_formats}
\end{table}

SWAP gate has many variation forms in practical implementations and we consider two specific cases in this work as shown in Tab. \ref{tab:swap_formats}. According to the linear superposition properties and linkage structures of the time evolution under study as shown in Fig. \ref{fig1:layout_sketch}, it suffices to examine the gate process of four specific initial states, namely the singlet and triplet states. For a `standard' format of SWAP gate as the left part in Tab. \ref{tab:swap_formats}, wave functions of all the four different initial states should end up with full population return, in which $\frac{1}{\sqrt{2}}(|01\rangle - |10\rangle)$ and $\frac{1}{\sqrt{2}}(|01\rangle + |10\rangle)$ should receive $\pi$ difference in phase. For an `opposite' format of SWAP gate as the right part in Tab. \ref{tab:swap_formats}, wave functions start in the initial state of $|00\rangle$ will end up in $|11\rangle$ and vice versa for that of $|11\rangle$, while $\frac{1}{\sqrt{2}}(|01\rangle - |10\rangle)$ and $\frac{1}{\sqrt{2}}(|01\rangle + |10\rangle)$ should receive no phase difference.

Previously, we can generally interpret the principles of the entangling phase gates \cite{OptEx480513, PhysRevApplied.20.L061002, Yuan2024FR2} as letting the qubit atoms' wave functions receive the correct state-dependent conditional phase shifts, which originate form dynamical phase, geometric phase or both. On the other hand, the Rydberg blockade SWAP gate inevitably involves net population transfer with respect to the computational basis $|00\rangle, |01\rangle, |10\rangle, |11\rangle$. This subtle difference will bring noticeable change in the design such as that now in the SWAP gate $|0\rangle, |1\rangle$ both experience the ground-Rydberg transition, which also causes non-trivial variation of the underlying dynamics.

With the above principles outlining the mechanisms of Rydberg blockade SWAP gate, the next question is to find out whether the desired optical driving exists and compatible with experimental reality. Analytically attaining solutions seems too difficult for now due to the nature of the problem as inverse solving. Alternatively, we seek viable results via numerical search or optimization like the development of Rydberg blockade entangling phase gates so far \cite{PhysRevApplied.13.024059, PhysRevApplied.20.L061002}. Without loss of generality, we set the criteria on the Rabi frequency waveforms that they start and end at intensity zero with first order derivative zero, for the ease of practicing modulation experimentally.

Here we choose Fourier series as the basis to express and generate the waveforms, which also naturally helps to eliminate or suppress high-frequency components \cite{PhysRevApplied.20.L061002}. More specifically, the function $f$ of Rabi frequency or detuning terms can be expressed as $[a_0, a_1, \ldots, a_N]$, representing $f(t)=2\pi \times \big(a_0 + \sum_{n=1}^{N} a_n\exp(2\pi i nt/\tau) + a^*_n\exp(-2\pi i nt/\tau) \big)/(2N+1) \text{ MHz}$ for a given reference time $\tau = 0.25\, \mu\text{s}$.

\section{Design appropriate waveforms}
\label{sec:results}

According to the discussions so far, we need to pay specific attention to the time evolutions of four initial states $|00\rangle, \frac{1}{2}(|01\rangle-|10\rangle), \frac{1}{2}(|01\rangle+|10\rangle), |11\rangle$, and in particular the wave functions associated with these initial states which we express as $\Psi_{|00\rangle}(t), \Psi_{\frac{1}{2}(|01\rangle-|10\rangle)}(t), \Psi_{\frac{1}{2}(|01\rangle+|10\rangle)}(t), \Psi_{|11\rangle}(t)$. It turns out that reasonable solutions of Rydberg blockade SWAP gate do exist after extensive search in the parameter space. On practical account, many numerical search algorithms can efficiently find appropriate values of $a_n$'s to satisfy the requirement of the SWAP gates under study. And not surprisingly, the solutions are not unique as the case in the previous studies of Rydberg blockade Controlled-PHASE gate. 

The Rydberg blockade effect provides the necessary interaction between two qubit atoms, enabling the exchange of quantum information embedded in the qubit atoms. As a natural consequence of Quantum No-Cloning, the quantum information does not receive duplication during the process. Meanwhile, the transition linkage structure plays a special role in making the Rydberg blockade SWAP gate possible. More interestingly, the results of our calculations suggest that the Rydberg dipole-dipole interaction strength on the order of $\sim 100$ MHz suffices for satisfying gate fidelity with respect to known experimental capabilities and from the practical point of view this indicates fast gate operation \cite{Yuan2024FR2}. 

\begin{figure}[h]
\centering
\includegraphics[width=0.432\textwidth]{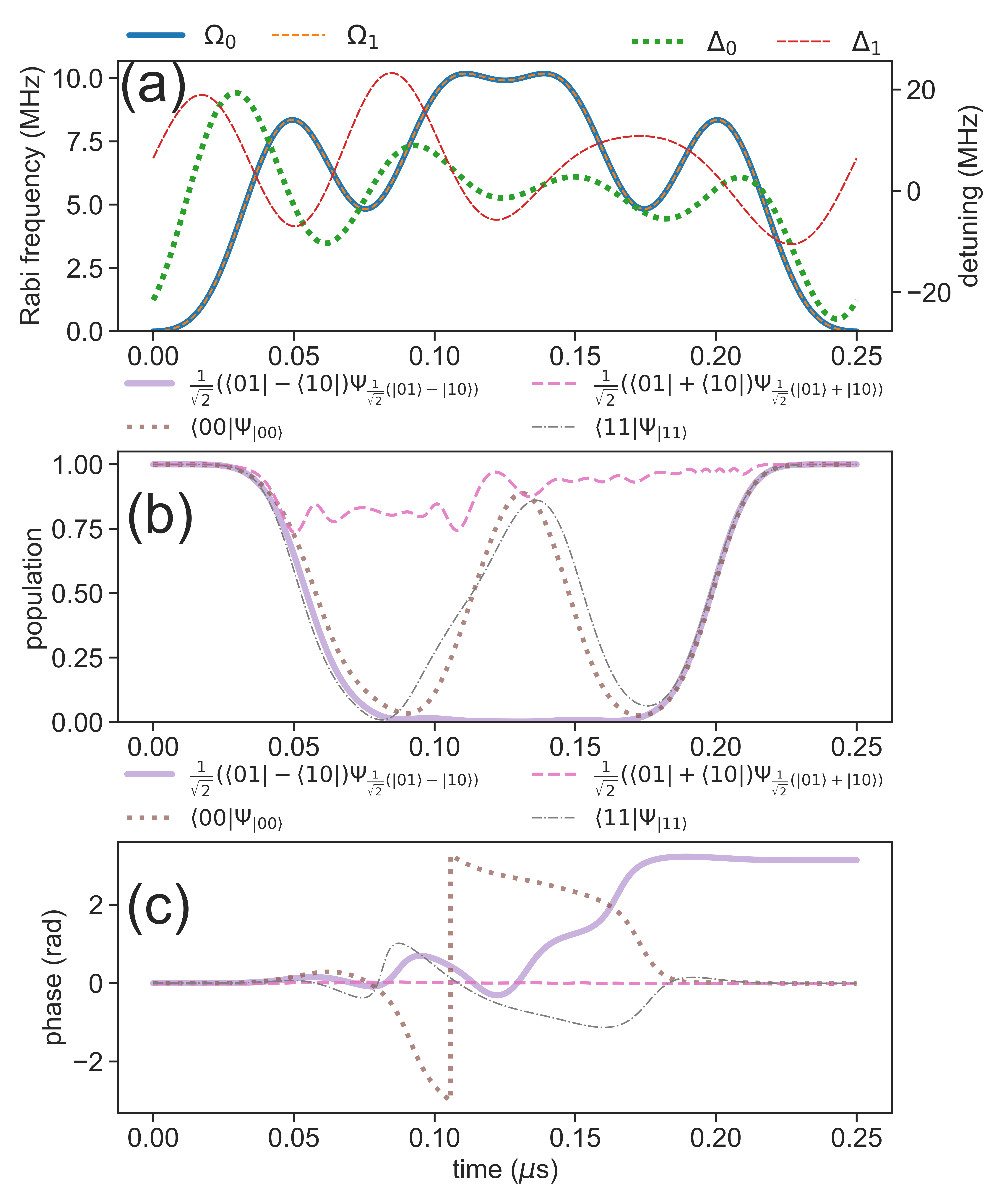}
\caption{(Color online) Numerical simulation of Rydberg Rydberg blockade SWAP gate with hybrid modulation. (a) Waveforms of modulation.  (b) Populations of wave functions corresponding to (a). (c) Phases of wave functions corresponding to (a). The calculated gate errors are less than $10^{-4}$.}
\label{fig2:ORMD_hybrid_modulation}
\end{figure}

We first look at the category with both amplitude and frequency modulations, namely the hybrid modulation. The Rabi frequency waveforms of the two driving lasers $\Omega_0(t), \Omega_1(t)$ can be proportional or even the same, providing certain degrees of extra conveniences to experimental implementation. Here, we let $\Omega_0=\Omega_1=$[66.26, -20.40, -4.41, -12.61, -1.47, 5.76], $\Delta_0=$[-13.37, -21.68-19.43$i$, -10.79-19.20$i$, -25.28-33.11$i$, -45.53-20.21$i$, -8.44+1.27$i$], $\Delta_1=$[54.53, -9.77-16.65$i$, -12.84-11.28$i$, 46.29-34.26$i$, -11.51-31.76$i$, -4.05+2.88$i$], and Fig. \ref{fig2:ORMD_hybrid_modulation} shows the waveforms and key features of the atomic wave functions during the time evolution. This result corresponds to a SWAP gate in the standard format as given by the left column in Tab. \ref{tab:swap_formats}. 

\begin{figure}[h]
\centering
\includegraphics[width=0.432\textwidth]{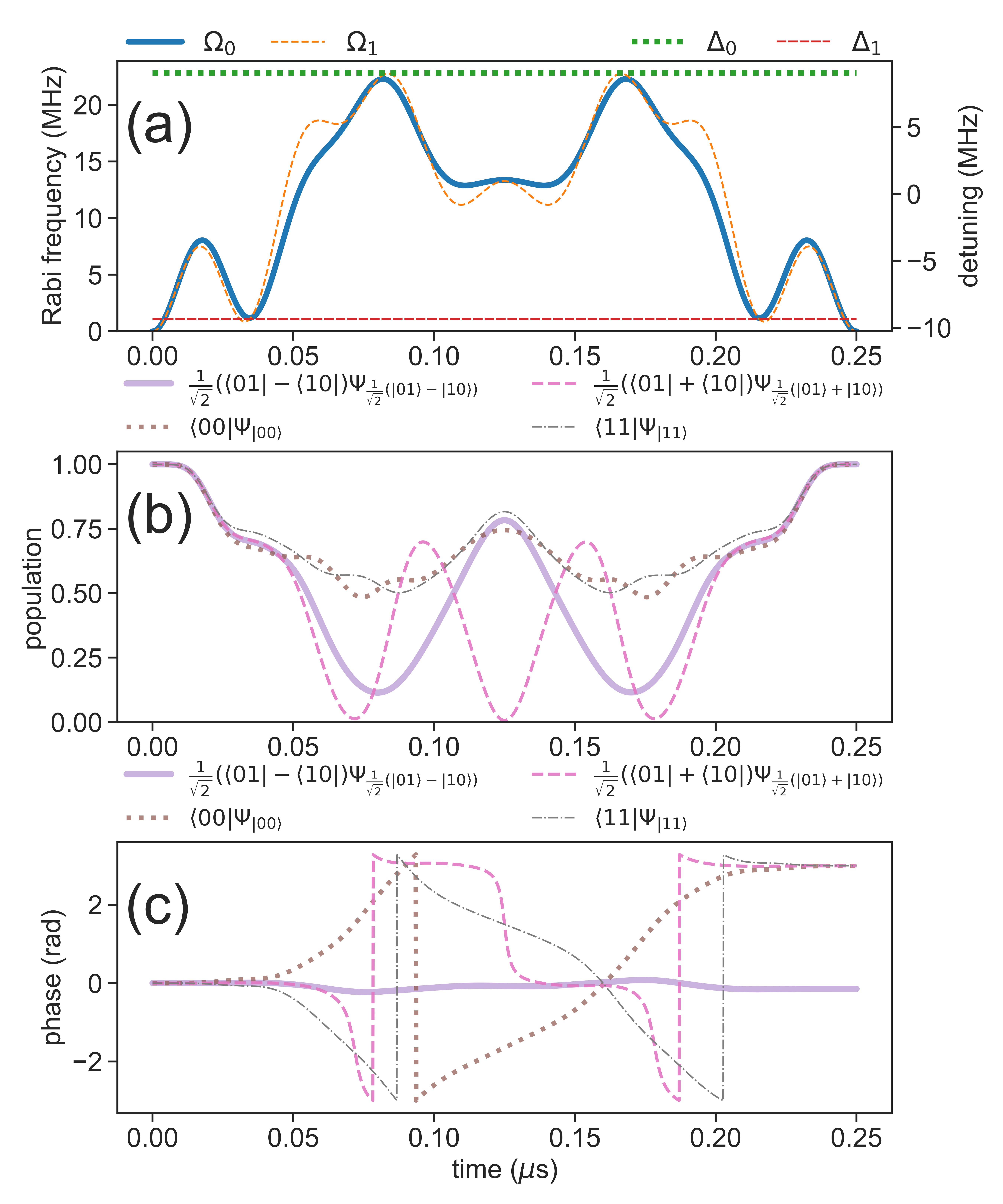}
\caption{(Color online) Numerical simulation of Rydberg Rydberg blockade SWAP gate with only amplitude modulation under off-resonant drivings. (a) Waveforms of modulation.  (b) Populations of wave functions corresponding to (a). (c) Phases of wave functions corresponding to (a). The calculated gate errors are less than $10^{-4}$.}
\label{fig3:ORMD_amplitude_modulation}
\end{figure}

If only amplitude modulation functions adequately for the task without involving the extra burden of frequency modulation, this will sometimes simplify the experimental implementation of Rydberg blockade gates. We investigate this possibility and obtain such results. In particular, we let $\Omega_0$=[199.45, -59.56, -34.98, 32.30, 6.86, -11.85, -6.17, -17.72, -8.61], $\Omega_1=$[206.49, -55.67, -48.16, 27.76, 11.52, -3.03, -2.06, -25.43, -8.19], $\Delta_0=[9.05]$, $\Delta_1=[-9.34]$, and Fig. \ref{fig3:ORMD_amplitude_modulation} shows the waveforms and key features of the atomic wave functions during the time evolution. Here, we particularly choose a result that does not yield the standard format of SWAP gate, but with an overall $\pi$ phase shift to demonstrate the versatility of our method of design.

While the above two typical categories of modulation styles suffice for the purpose of experimental realizations, we want to emphasize that even for the case of single-photon ground-Rydberg transition there exists more degrees of freedom in the method of synthetic continuously-modulated driving. For example, we can apply only amplitude modulation with resonant conditions of $\omega_1, \omega_2$, where this and extra examples will be provided in the appendix. Furthermore, this category of gate process allows the design of many variation forms of SWAP gate, which basically reduce to determining appropriate parameters of the optical driving.

\section{Estimation of performance against common adverse effects}

While the abstract concept and computational formulation of Rydberg blockade SWAP gate seem plausible in the above analysis, its potential performance under realistic experimental conditions needs further consideration and examination. Here we proceed to estimate and discuss this subject with respect to a few typical and commonly encountered adverse effects. One key question is whether the gate process still holds at a reasonable quality against certain degrees of deviations from the idealized theoretical condition. Fortunately and interestingly, the synthetic continuously-modulated driving, when employed in the Rydberg blockade SWAP gate, also implicitly triggers the two-atom dark state under the presence of Rydberg blockade effect \cite{PhysRevA.96.042306}, which offers substantial help to the robustness of gate. Meanwhile, we also hope to identify and analyze the major difficulties if gate error on the order of $10^{-4}$ or smaller is desired. If the experiment works with the two-photon ground-Rydberg transition instead of the single-photon transition, then one needs to pay extra care in suppressing the population of the intermediate level which usually has a much larger decay rate than the ground and Rydberg levels.  

\begin{figure}[h]
\centering
\begin{minipage}{0.233\textwidth}
	\centering
    \includegraphics[width=\textwidth]{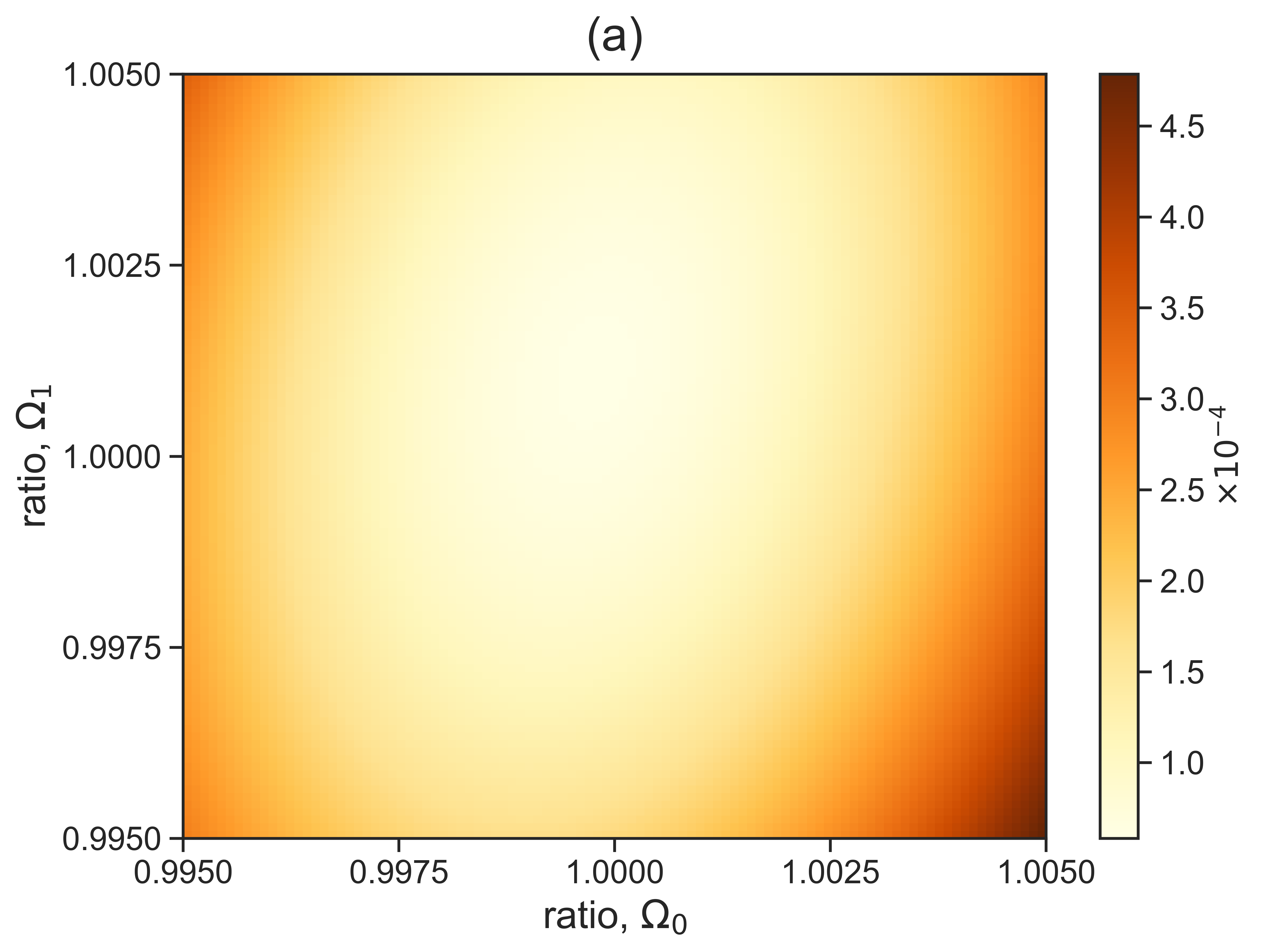}    
\end{minipage}
\hfill
\begin{minipage}{0.233\textwidth}
	\centering
    \includegraphics[width=\textwidth]{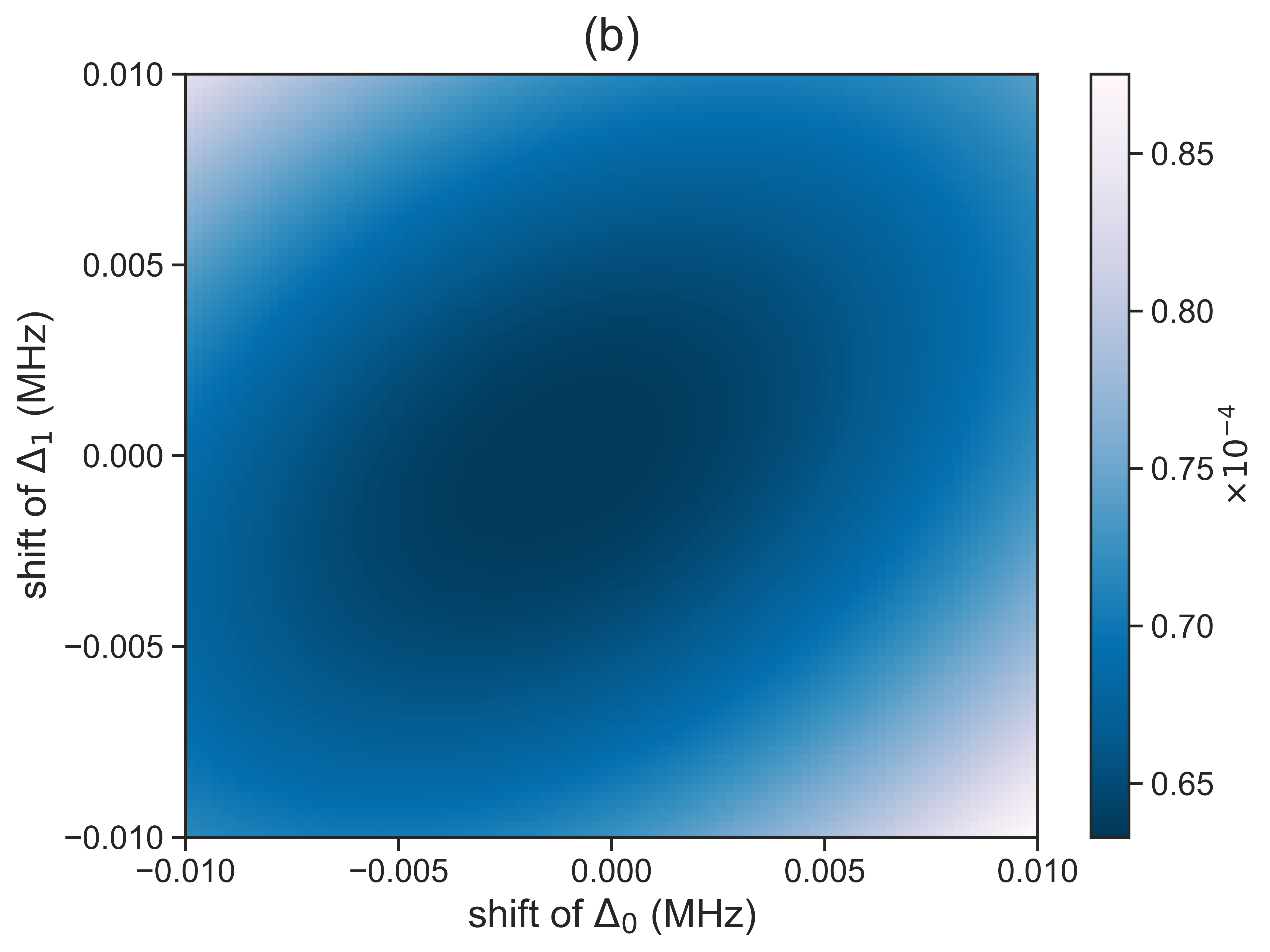}    
\end{minipage}
\caption{(Color online) Numerical simulation of gate errors with respect to changes of Rabi frequencies and detunings. (a) Varying ratios of $\Omega_0, \Omega_1$. (b) Adding constant shifts to the detunings.}
\label{fig4:scan_Rabi+detuning}
\end{figure}

First of all, we would like to calculate the influences of possible errors in the Rabi frequencies and detunings of driving laser fields. This type of adverse effects occur commonly in typical cold atom qubit experiments, and it makes sense to require the SWAP gate to maintain reasonable performance as long as these errors stay within a technically possible range. We show a typical result in Fig. \ref{fig4:scan_Rabi+detuning} by studying the gate errors of waveforms in Fig. \ref{fig2:ORMD_hybrid_modulation} under these adverse effects. It suggests that the experimental calibration of Rabi frequency needs to reach certain precision for a high-fidelity gate. 

\begin{figure}[h]
\centering
\begin{minipage}{0.233\textwidth}
	\centering
    \includegraphics[width=\textwidth]{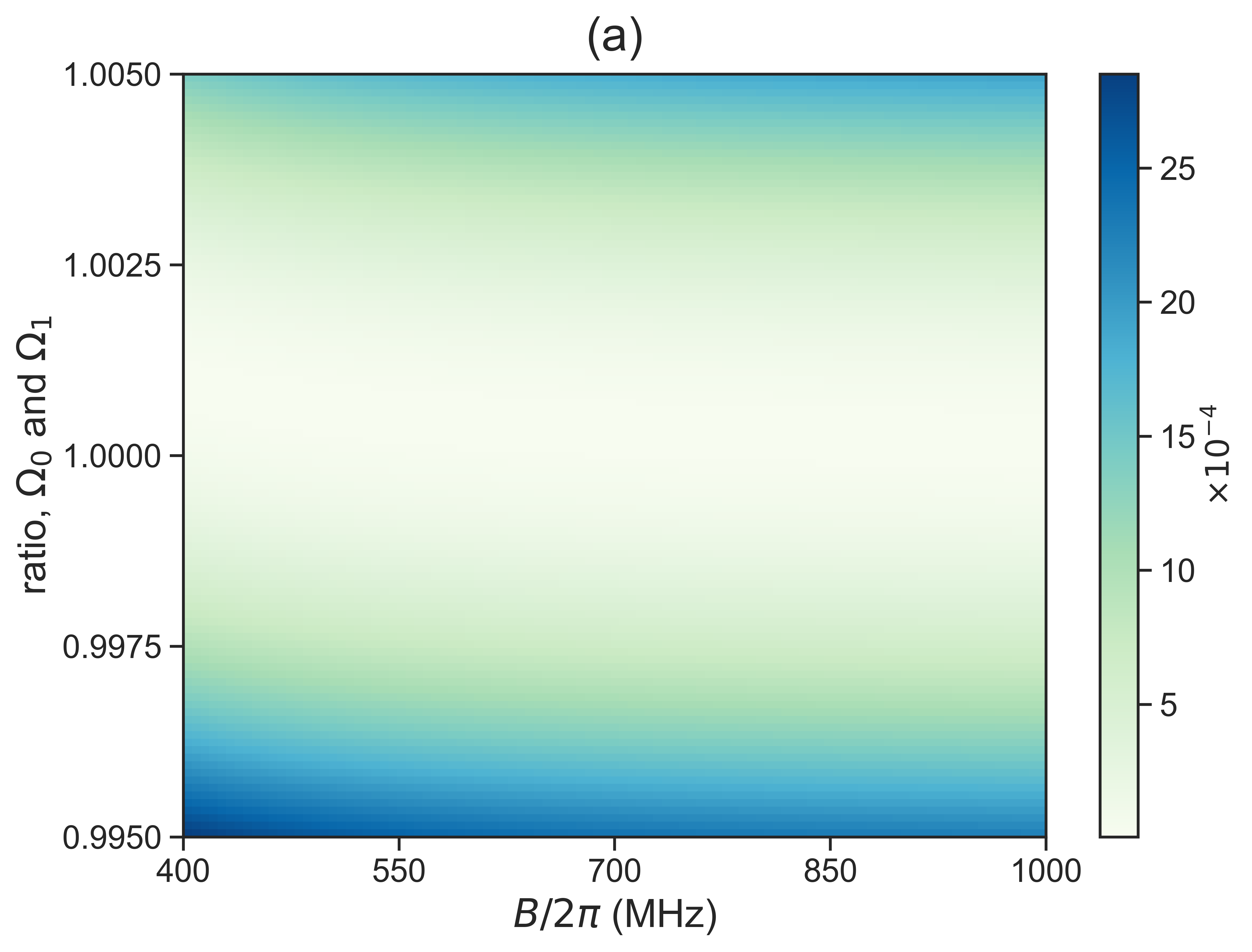}    
\end{minipage}
\hfill
\begin{minipage}{0.233\textwidth}
	\centering
    \includegraphics[width=\textwidth]{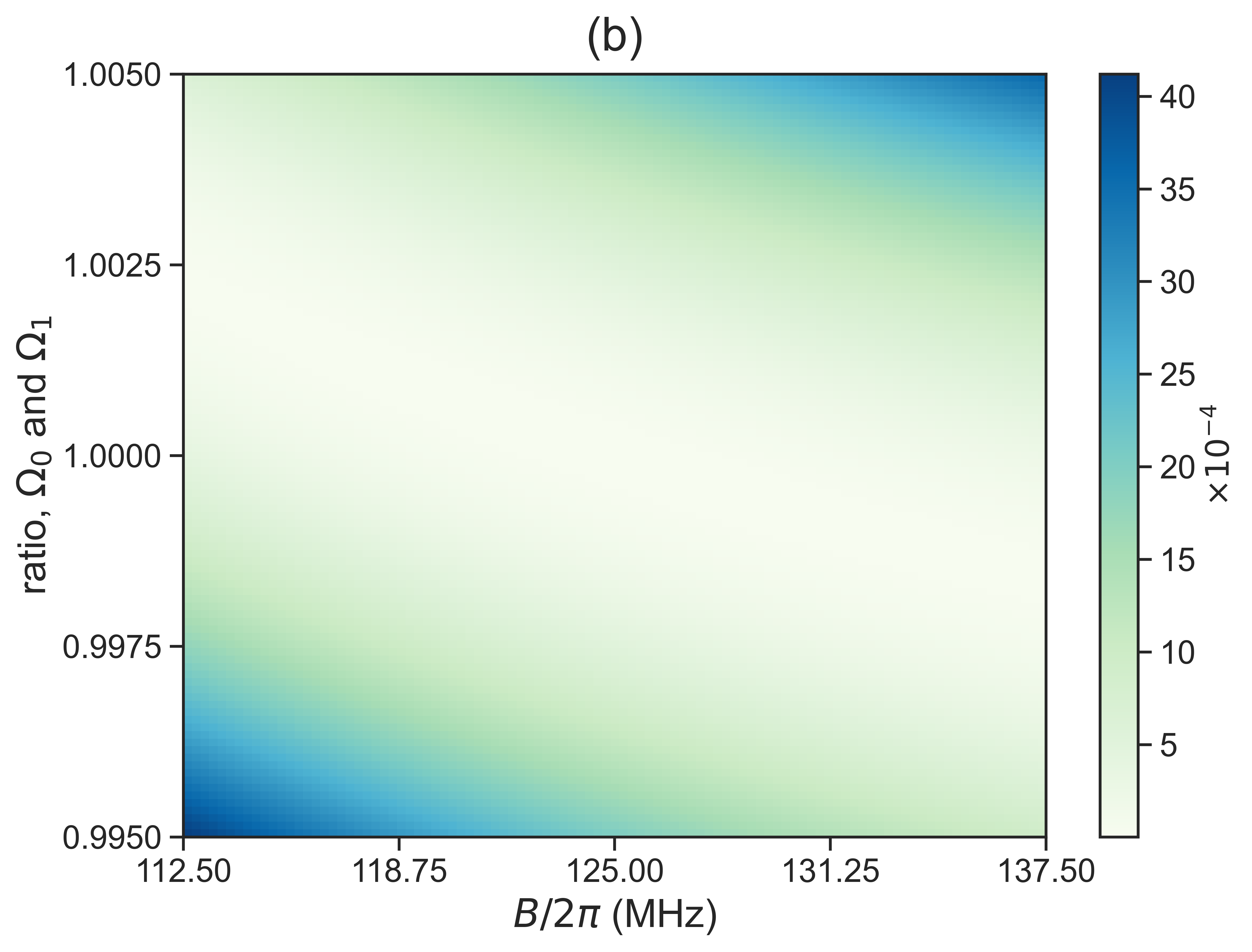}    
\end{minipage}
\caption{(Color online) Numerical simulation of gate errors with respect to changes of Rydberg dipole-dipole interaction strength. (a) Estimated performance of waveforms designed for $B=\infty$. (b) Estimated performance of waveforms designed for $B=2\pi\times 125$ MHz.}
\label{fig4:scan_forster_strength}
\end{figure}

Generally speaking, when designing cold atom qubit arrays, we hope to accurately fix the positions of qubit atoms all across the board and henceforth the distances between qubit atoms as well. Nevertheless, sometimes the deviations of the inter-atomic distances from design can become significant and relatively difficult to accurately track. This easily causes the varying of Rydberg dipole-dipole interaction strength, and sometimes the Rabi frequencies seen by the atoms too. The waveforms in Fig. \ref{fig3:ORMD_amplitude_modulation} are designed for idealized Rydberg blockade effect, and on its basis we further develop a variation form specifically tailored for finite Rydberg blockade strength $B=2\pi\times 125$ MHz, $\delta_q=0$. And then we proceed to examine the influences of varying the overall ratio of Rabi frequencies $\Omega_0, \Omega_1$ and the Rydberg blockade strength $B$ to emulate the experimental adverse effects. The numerical result presented in Fig. \ref{fig4:scan_forster_strength} indicates the inherent advantage of Rydberg blockade gate that remains relatively high quality performance a over a reasonable range of $B$. We observe that the Rydberg blockade SWAP gate here also has the versatility to adapt to the actual condition of Rydberg dipole-dipole interaction, similar to the case of Controlled-PHASE gate \cite{OptEx480513}.

Within the scope of our discussions, we assume uniform laser drivings such that $|\Omega_0|, |\Omega_1|$ are the same on both qubit atoms, and this corresponds to the preferred experimental style especially when executing parallel operation on many atom pairs. Even if the control and target qubit atoms have different driving lasers, the design of Rydberg blockade SWAP gate still applies. In particular, for the case that control and target qubit atoms have different driving Rabi frequencies, the singlet type and the triplet type linkage structures as in Fig. \ref{fig1:layout_sketch} do not necessarily remain, but the method of synthetic continuously-modulated driving can still help to find the appropriate SWAP gate waveform if it exist. On the other hand, the spontaneous emission of Rydberg levels imposes a fundamental limit on the practical gate fidelity. For the Rydberg blockade SWAP gate time $T$ and Rydberg level decay rate $\gamma_r$, it causes a gate error on the order of $\sim 0.5\gamma_rT$, as in Rydberg blockade Controlled-PHASE gates extensive studied previously. If other experimental conditions allow, a faster gate seems as the future direction as long as the higher order frequency components in the waveforms have been handled carefully \cite{PhysRevApplied.20.L061002}.

The recent innovations of buffer-atom-mediated quantum logic gates and buffer atom framework \cite{Yuan2024SCPMA, Yuan2024FR2} aim at enhancing connectivity and suppressing cross-talk without mechanically shuttling the qubit atoms. Our Rydberg blockade SWAP gate is readily compatible with the buffer atom framework at the cost of extra lasers and extra control systems. In particular, we can apply the Rydberg blockade SWAP gate between qubit and buffer atoms such that the buffer atom or buffer atom relay can effectively help to operate SWAP gate between qubit atoms, where buffer and qubit atoms different elements so that cross-talk does not become an issue. When the two driving lasers $\omega_0, \omega_1$ are parallel and co-propagating, the net atom-photon momentum exchange reduces to zero on average and henceforth the Rydberg blockade SWAP gate with synthetic continuously-modulated driving does not cause heating. On the other hand, if the qubit atoms are heated up during computational quantum logic gate operations for other reasons, the Rydberg blockade SWAP gate with a nearby buffer atom provides a tool to cool down the qubit atoms in the run. With these analysis, we anticipate the future large-scale cold atom qubit array can maintain fast gate speed, high connectivity, and high fidelity simultaneously by combining the buffer atom framework and Rydberg blockade SWAP gate with synthetic continuously-modulated driving.

\section{Conclusion and Outlook}

In conclusion, we design and analyze the fast SWAP Rydberg blockade gates with synthetic continuously-modulated driving. This category of SWAP gates can work with one-photon or two-photon ground-Rydberg transitions, and is readily compatible with the currently available hardwares in the research field of cold atom qubits. According to the results of Rydberg blockade Controlled-PHASE gate, synthetic continuously-modulated driving leads to competitive performance in practical implementation, and here we observe that the Rydberg blockade SWAP gate inherits many of the previously known advantages and features of Controlled-PHASE gate. With the BAM gate mediating the connection between non-interacting qubit atoms, we anticipate that the combination of fast Rydberg blockade SWAP gate and BAM gate will enhance the connectivity of very-large-scale cold atom qubit array in the near future, and such that an entangling gate between two relatively distant qubit atoms can operate with high speed and high fidelity.
Whilst a fast and high-fidelity SWAP gate is known as an indispensable ingredient of superconducting quantum processors, a lot of recent findings together with our results hint that the superconducting qubit platform and the neutral atom qubit platform will find more and more common ground. In the future, we are looking forward to that many of the previously established quantum coherent control methods, quantum algorithms and quantum error correction codes of superconducting qubits can extend to the neutral atom qubits.

Xin Wang and Tianze Sheng contributed equally to this work. The authors gratefully acknowledge the support from the National Natural Science Foundation of China (Grant No. 92165107) and the Science and Technology Commission of Shanghai Municipality.

\appendix

\section{Extra details and more examples}

\begin{figure}[h]
\centering
\includegraphics[width=0.46\textwidth]{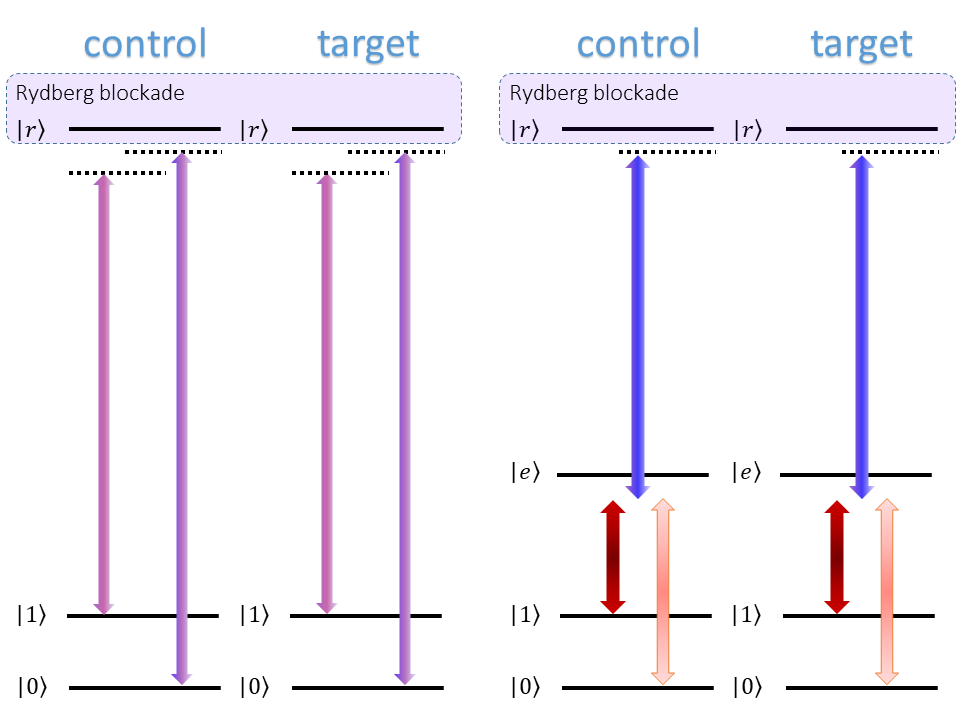}
\caption{(Color online) Comparison of Rydberg blockade SWAP gates via one-photon ground-Rydberg transition (on the left) and two-photon transition (on the right).}
\label{figa1:2photon_transition}
\end{figure}

In the main text, we have focused the discussions on the scenario of single-photon ground-Rydberg transition. In fact, the concept and method of Rydberg blockade gate with synthetic continuously-modulated driving applies to both the single-photon and two-photon ground Rydberg transitions, where the two-photon transition allows a rich variety of modulation styles. We present a typical driving pattern for two-photon ground-Rydberg transition here in Fig. \ref{figa1:2photon_transition}. The single-photon transition waveforms can be translated to the two-photon transition waveforms directly by the technique of adiabatic elimination. Further possibilities include only modulate one driving laser out of the two, just like the previous instantiations in the Rydberg entangling phase gates \cite{PhysRevA.105.042430, OptEx480513}.

\begin{table}[h]
\begin{center}
\Scale[1.0]{  
  \begin{tabular}{| c | c | c |}  
    \hline
    $C_{1-}$ & $C_{0r}$ & $C_{1r}$ \\ \hline
    $\frac{|01\rangle-|10\rangle}{\sqrt{2}}$ & $\frac{|0r\rangle-|r0\rangle}{\sqrt{2}}$ & $\frac{|r1\rangle-|1r\rangle}{\sqrt{2}}$\\ \hline
  \end{tabular}
  }
\end{center}
\caption[Labeling singlet states]{Labeling of basis states associated with the dynamics of $C_\text{singlet}$.}
\label{tab:singlet_states_names}
\end{table}

\begin{table}[h]
\begin{center}
\Scale[1.0]{
  \begin{tabular}{ | c | c | c | c | c | c| c |}  
    \hline
    $C_0$ & $C_{1+}$ & $C_2$ & $C_{r0}$ & $C_{r1}$ & $C_{rr}$ & $C_{qq'}$ \\ \hline
    $|00\rangle$ & $\frac{|01\rangle+|10\rangle}{\sqrt{2}}$ & $|11\rangle$ & $\frac{|0r\rangle+|r0\rangle}{\sqrt{2}}$ & $\frac{|r1\rangle+|1r\rangle}{\sqrt{2}}$ & $|rr\rangle$ & $|qq'\rangle$\\ \hline
  \end{tabular}   
  }
\end{center}
\caption[Labeling triplet states]{Labeling of states associated with the dynamics of $C_\text{triplet}$.}
\label{tab:triplet_states_names}
\end{table}

Next, we would like to discuss the details of the Schr\"odinger equations for the relevant gate processes as shown in Fig. \ref{fig1:layout_sketch}. While we have already explained the Hamiltonian of system under study in the main text, we seek here to compute the the time evolution of wave functions $C_\text{singlet}, C_\text{triplet}$, where $C_\text{singlet}=[C_{1-}, C_{0r}, C_{1r}]$ and  $C_\text{triplet}=[C_0, C_{1+}, C_2, C_{r0}, C_{r1}, C_{rr}, C_{qq'}]$. Due to the inherent symmetry of the system, the time evolutions of $C_\text{singlet}$ and $C_\text{triplet}$ separate from each other and have mutually independent Schr\"odinger equations. The notations of these wave functions are shown in Tab. \ref{tab:singlet_states_names} and Tab. \ref{tab:triplet_states_names}. 

We adhere to the Schr\"odinger picture and the equations of motion for the wave functions include: $i\hbar\frac{d}{dt} C_\text{singlet} = H_\text{singlet}(t) C_\text{singlet}$ and $i\hbar\frac{d}{dt} C_\text{triplet} = H_\text{triplet}(t) C_\text{triplet}$. The equations associated with $C_\text{singlet}$ look similar to the typical $V$-type three-level system as follows. Here, the rotating wave frame and rotating wave approximation are carried out in the usual way as other Rydberg atom analysis with the rotating frequency chosen as the driving laser frequency.

\begin{equation}
\label{eq:H_singlet}
H_\text{singlet} = \frac{\hbar}{2}
\begin{bmatrix}
0 & \Omega_0 & \Omega_1 \\
\Omega^*_0 & 2\Delta_0 & 0 \\
\Omega^*_1 & 0 & 2\Delta_1
\end{bmatrix}
\end{equation}

On the other hand, it needs a little extra care to handle the rotating wave frame's phase terms in expressing $H_\text{triplet}$ because of some states receive the couplings from more than one driving laser. More specifically, here we note that no extra rotating phases are imposed on $|00\rangle$, $\frac{|01\rangle+|10\rangle}{\sqrt{2}}$ and $|11\rangle$. Then the equations of motion for $C_\text{triplet}$ are given in the following Eq. \eqref{eq:H_triplet}.

\begin{widetext}

\begin{equation}
\label{eq:H_triplet}
H_\text{triplet} = \frac{\hbar}{2}
\begin{bmatrix}
0 & 0 & 0 & \sqrt{2}\Omega_0 & 0 & 0 & 0 \\
0 & 0 & 0 & \Omega_1 e^{i(\Delta_0 - \Delta_1)t} & \Omega_0 e^{i(\Delta_1 - \Delta_0)t} & 0 & 0 \\
0 & 0 & 0 & 0 & \sqrt{2}\Omega_1 & 0 & 0 \\
\sqrt{2}\Omega^*_0 & \Omega^*_1e^{i(\Delta_1-\Delta_0)t} & 0 & 2\Delta_0 & 0 & \sqrt{2}\Omega_0 e^{i(\Delta_1-\Delta_0)t} & 0 \\
0 & \Omega^*_0e^{i(\Delta_0-\Delta_1)t} & \sqrt{2}\Omega^*_1 & 0 & 2\Delta_1 & \sqrt{2}\Omega_1e^{i(\Delta_0-\Delta_1)t} & 0\\
0 & 0 & 0 & \sqrt{2}\Omega^*_0 e^{i(\Delta_0-\Delta_1)t} & \sqrt{2}\Omega^*_1e^{i(\Delta_1-\Delta_0)t} & 2(\Delta_0+\Delta_1) & 2B\\
0 & 0 & 0 & 0 & 0 & 2B^* & 2(\Delta_0 + \Delta_1 + \delta_q)
\end{bmatrix}
\end{equation}

\end{widetext}

We have already demonstrated typical and generic cases of Rydberg blockade SWAP gate waveforms for various specific conditions in the main text. In particular, we have mentioned the design of waveforms to adapt to finite Rydberg blockade strength. There exists straightforward design process to achieve this functionality. As we can observe from Eq. \eqref{eq:H_singlet} and Eq. \eqref{eq:H_triplet}, only the triplet linkage system involves the Rydberg dipole-dipole interaction whose model is given by $B, \delta_q$ here. In the design process, we let $B \to \infty$ for idealized Rydberg blockade effect and let $B, \delta_q$ be some prescribed fixed values for the case of finite blockade strength. According to the discussions in the main text, even if the design of gate waveforms aims specifically for a particular Rydberg blockade strength, its performance will look reasonable over some range. This comes from the natural advantage of the Rydberg blockade gate that not requiring a precise tuning of the Rydberg dipole-dipole interaction. For instance, in the main text we mentioned a variation form of waveforms specifically tailored for finite Rydberg blockade strength $B=2\pi\times 125$ MHz, $\delta_q=0$, whose details are: $\Omega_0=$[201.72, -60.68, -35.18, 33.33, 6.14, -11.66, -6.12, -18.53, -8.16], $\Omega_1=$[208.18, -57.17, -47.92, 28.42, 11.19, -3.41, -1.91, -24.41, -8.89], $\Delta_0=[9.07]$, $\Delta_1=[-9.38]$.

\begin{figure}[h!]
\centering
\includegraphics[width=0.432\textwidth]{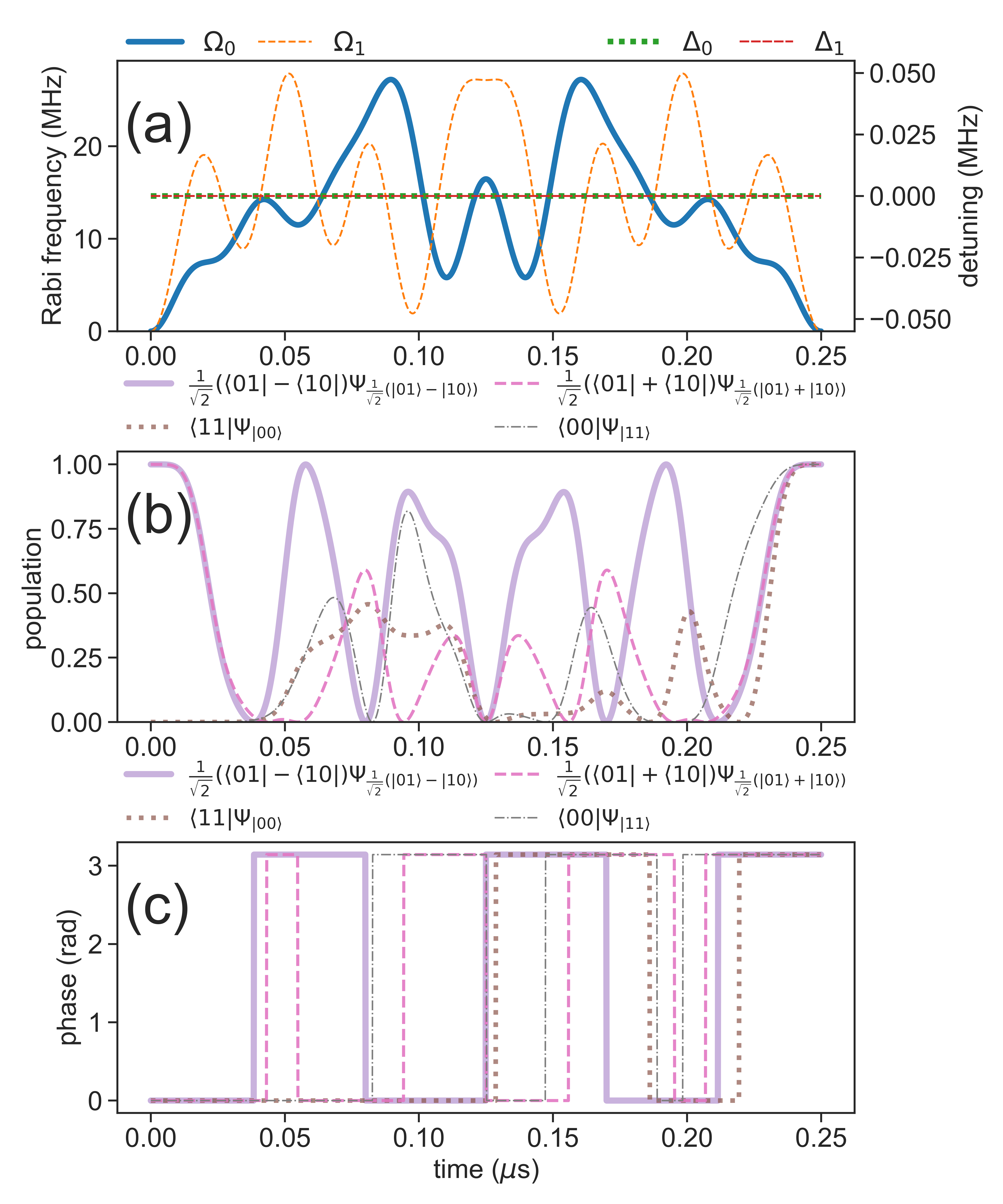}
\caption{(Color online) Numerical simulation of Rydberg Rydberg blockade SWAP gate with only amplitude modulation under almost-resonant drivings. (a) Waveforms of modulation. (b) Populations of wave functions corresponding to (a). (c) Phases of wave functions corresponding to (a). The calculated gate errors are less than $10^{-4}$.}
\label{figa2:ARMD_amplitude_modulation}
\end{figure} 

\begin{figure}[h]
\centering
\includegraphics[width=0.432\textwidth]{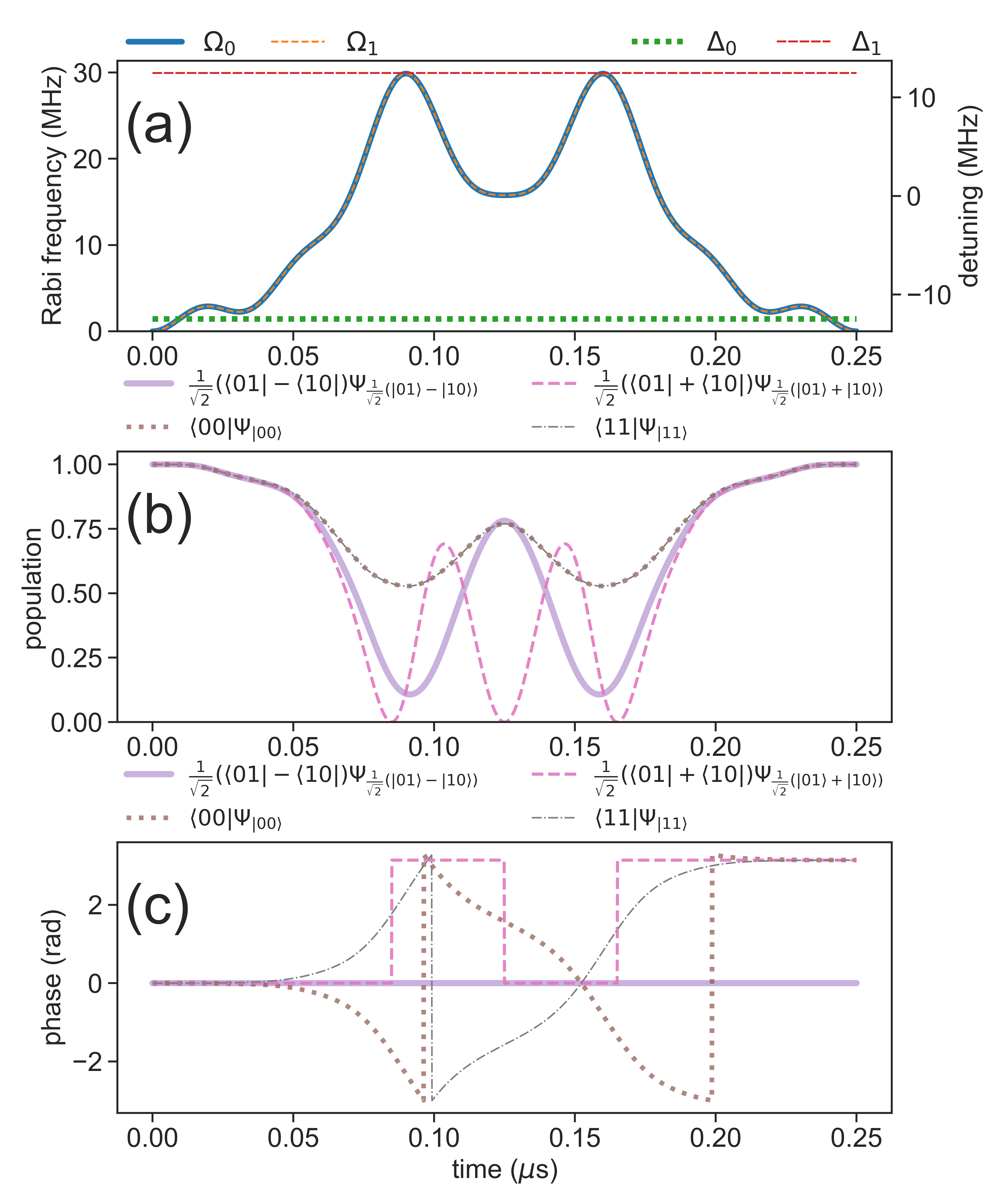}
\caption{(Color online) Numerical simulation of Rydberg Rydberg blockade SWAP gate with only amplitude modulation under off-resonant drivings. (a) Waveforms of modulation, in particular, the Rabi frequency amplitudes of $\Omega_0, \Omega_1$ are identical. (b) Populations of wave functions corresponding to (a). (c) Phases of wave functions corresponding to (a). The calculated gate errors are less than $10^{-4}$.}
\label{figa3:ORMD_amplitude_modulation}
\end{figure}

In order to provide more insights into the method of synthetic continuously-modulated driving, we discuss a few more examples here as in Fig. \ref{figa2:ARMD_amplitude_modulation} and Fig. \ref{figa3:ORMD_amplitude_modulation}, with some extra details or characteristics omitted in the main text. 

Fig. \ref{figa2:ARMD_amplitude_modulation} corresponds to the concept of quantum gates via almost-resonant driving. In particular, here we have $\Delta_0=\Delta_1=0$ such that the two driving lasers $\omega_0, \omega_1$ theoretically hit the resonances of their transitions respectively. More specifically, here we have $\Omega_0$=[250.95, -51.38, -51.18, 25.85, -32.11, -12.74, 24.64, -21.83, 11.38, -18.10] and $\Omega_1$=[286.47, -20.74, -13.46, -55.45, 30.51, -27.91, 12.23, -49.65, -43.50, 24.74]. As the successful outcome of SWAP gate requires both population transfer and phase accumulation, we observe that the almost-resonant driving method yields an obviously different style of dynamics. For instance, the phases of wave functions always ideally jump between $0, \pi$ during the time evolution. Apparently, the source of these accumulated phases comes from the dynamical phase according to the analysis of quantum geometry \cite{BerryPhasePaper, PhysRevLett.58.1593, PhysRevLett.131.240001}.

Fig. \ref{figa3:ORMD_amplitude_modulation} shows another set of waveforms designed for finite Rydberg blockade strength with $B=2\pi\times 100$ MHz and $\delta_q=0$, under the interesting condition of $\Delta_0+\Delta_1=0$ exactly. Its details include $\Omega_0=$[191.04, -89.66, -18.38, 37.37, -22.13, 4.32, 4.10, -11.15], $\Omega_1=$[191.04, -89.66, -18.38, 37.37, -22.13, 4.32, 4.10, -11.15], $\Delta_0=[-12.48]$, $\Delta_1=[12.48]$. The special point here is that $\Omega_0(t) = \Omega_1(t)$, which can possibly bring some extra conveniences in experimental implementation.

\bibliographystyle{apsrev4-2}
\bibliography{maracas_ref}

\end{document}